\documentclass{article}

\usepackage{arxiv}

\usepackage[version=3]{mhchem} 
\usepackage{graphicx}
\usepackage{makecell}
\usepackage{tikz}
\usepackage{pgfplots}
\pgfplotsset{width=9cm,height=7cm,compat=1.3}
\usepackage{pgfplotstable}

\title{NMR shift prediction from small data quantities}

\author{Herman Rull\\
Department of Computer Science, Tartu University, Tartu, Estonia\\
\texttt{herman.rull@ut.ee}
\And
Markus Fischer\\
Institute for Medical Physics and Biophysics, Leipzig University, Germany\\
\And
Stefan Kuhn\\
Department of Computer Science, Tartu University, Tartu, Estonia\\
}

\begin{document}
\maketitle

\begin{abstract}
Prediction of chemical shift in NMR using machine learning methods is typically done with the maximum amount of data available to achieve the best results. In some cases, such large amounts of data are not available, e.g. for heteronuclei. We demonstrate a novel machine learning model which is able to achieve good results with comparatively low amounts of data. We show this by predicting $^{19}F$ and $^{13}C$ NMR chemical shifts of small molecules in specific solvents. 
\end{abstract}

\section{Introduction}

Prediction of chemical shift in nuclear magnetic resonance (NMR) is a long-standing problem in chemoinformatics. \cite{shoolery} is perhaps the earliest publication in the field. We define prediction here as methods using existing data as opposed to ab-initio calculations. Over time, various methods for such predictions have been developed. In particular, machine learning methods have been applied, starting with early methods, like small neural networks \cite{Kvasnicka1992}, up to the latest developments in convolutional and graph neural networks. We refer the reader to the recent review \cite{PMID:34787335} for an overview.

For supervised learning methods, like the mentioned neural networks, annotated datasets are needed, and the number of data points used is a significant factor in the quality of the predictions. A review of the literature shows, that the datasets used are generally big, consisting of tens of thousands of molecules. Table~\ref{table:litnumbers} gives an overview of the number of structures used in recent publications. It should be noted, that sometimes preliminary selection was employed, e.g. the 17,000 structures of \,\cite{doi:10.1021/acs.orglett.2c01251} are a selection (using Morgan fingerprints and the MaxMin algorithm from RDKit) from 170,000 molecules by structural diversity.

\begin{table*}[h]
\centering
\begin{tabular}{ c | c c c c c c c c } 
\hline
Literature reference & \makecell{\cite{pmid31388784}} & \makecell{\cite{pmid33428408}} & $^{13}C$ \cite{doi:10.1021/acs.jcim.0c00195} & $^{1}H$ \cite{doi:10.1021/acs.jcim.0c00195} & \makecell{\cite{C9SC03854J}} & \makecell{\cite{mestrelab}} & \makecell{\cite{D1SC03343C}} & \makecell{\cite{doi:10.1021/acs.orglett.2c01251}}\\
\hline
\makecell{number of\\structures used} & 32,538 & 57,456 & 21,481 & 10,248 & 75,382 & 400,000 & 8000 & 17,000\\
\hline
\end{tabular}
\caption{Examples of papers about chemical shift prediction and the number of structures used.}
\label{table:litnumbers}
\end{table*}

Unfortunately, many studies do not take  the influence of the amount of training data on the quality of the prediction into account. An exception is \,\cite{9842465}\,, which shows that some machine learning methods only show suitable predictive power with more than 5000 training examples. However, in many practical applications, the amount of experimental data available is quite limited. Examples are NMR chemical shifts of heteronuclei, specific classes of compounds, NMR spectra measured with particular solvents, or other certain experimental conditions. For such cases of small amounts of data (defined as less than 5000 structures here), the existing models either do not provide a good solution or are yet untested on such small datasets.

In this paper, we present a graph neural network that is able to achieve good predictions with small amounts of data and is competitive with state-of-the-art models for larger amounts of data. We restrict our research to small organic molecules in solution. The prediction of solid-state NMR chemical shifts, biological macromolecules, or inorganic compounds requires different, and more specific models. Therefore, we exclude those cases from the current research.

\section{Materials and Methods}

Predictive models can be applied to many different NMR parameters, such as coupling constants, relaxation time, or peak shape. Here, we want to focus on the prediction of NMR chemical shifts. Modern (solution) NMR experiments are an excellent source of data, as they provide isotropic chemical shift information with little noise\cite{ZANGGER20151}. In NMR spectroscopy, the chemical shift, which is equivalent to the resonance frequency of an atom, is determined by the (chemical) environment of a nucleus. The representation of this environment is a hard task. A single atom might be represented by a vector, in which its properties are stored. However, that representation won't work for an entire molecule. Because of this, we disregard tensor-like representation and constitute the molecule as a graph instead. There, the atoms can be described by nodes, and the connecting bonds can be described by edges. Geometric learning with graph neural networks \cite{https://doi.org/10.48550/arxiv.1806.01261,1555942} provides a suitable tool to run ML algorithms on such specific structures. In the graph, information is passed along the edges, making information from connected atoms the most important information. This is equivalent to real molecules, where neighbouring atoms, which are connected with a low number of bonds, have the most impact on NMR chemical shift values.

\begin{figure}
  \centering
  \includegraphics[width=10cm]{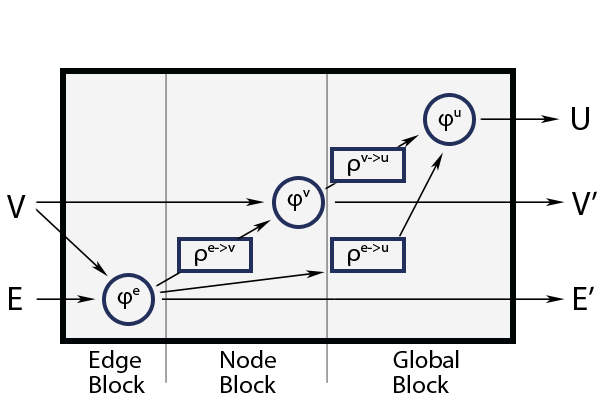}
  \caption{Information flow in the message-passing graph network of the 2023 model. In addition to node and edge aggregation functions, there is also an additional edge aggregation function feeding into the global update function (from \cite{FISCHER2022107750}).}
  \label{fig:messagepassng}
\end{figure}

For this work, we developed a model that learns the atomic properties in molecules, based on \,\cite{FISCHER2022107750}\,, which we call the ``2023 model'' in this paper. The model uses message-passing graph networks\cite{https://doi.org/10.48550/arxiv.1509.09292,https://doi.org/10.48550/arxiv.1704.01212}, which pass information via edges in the graph, in this way building up information locally in the nodes. Following \,\cite{FISCHER2022107750}\,, we use a type of message-passing graph network block with an additional edge aggregation function, shown in Figure~\ref{fig:messagepassng}. Since we employ the network to predict node-level features of small molecules, we disregard the second stage of the graph network described in \,\cite{FISCHER2022107750}\,. We use a set of features which is given in Tables~\ref{table:featuresatom} and \ref{table:featuresbond} to describe atoms and bonds. 

\begin{table*}[t]
\centering
\begin{tabular}{ c c } 
\hline
Feature & Description [unit] (type) \\
\hline
\makecell{atomic number} & \makecell{one hot encoded [all atoms in dataset](array[bool])}\\
atomic radius & Slater data from Mendeleev library [pm](int)\\
neutrons & number of neutrons [-](int) \\
electronegativity &  Pauling scale [-](float) \\
\makecell{electron affinity} & \makecell{value from Mendeleev library [eV](float)}\\
\hline
\end{tabular}
\caption{Atom features used in the 2023 model.}
\label{table:featuresatom}
\end{table*}

\begin{table*}[t]
\centering
\begin{tabular}{ c c } 
\hline
Feature & Description [unit] (type) \\
\hline
\makecell{bond length} & \makecell{distance between atom centers [\AA](float)}\\
\makecell{bond type} & \makecell{one hot encoded[single, double, triple, aromatic](array[bool])}\\
\hline
\end{tabular}
\caption{Bond features used in the 2023 model.}
\label{table:featuresbond}
\end{table*}

This specific feature selection was chosen because of preliminary experiments, which measured the impact of each feature on the final prediction. After that, the best-performing features were combined, until the prediction quality was no longer improving.

To optimize the prediction accuracy, we carefully selected the best hyperparameters for the 2023 model. The most important hyperparameters were the number of message-passing steps, the learning rate and the weight decay. The hyperparameters were optimized on the training set of $^{19} F$ data using 4-fold cross-validation. The best performing model used 6 message-passing steps, a learning rate of $10^{-3}$, and a weight decay of $0.01$.

All programming is done in Python using RDKit \cite{rdkit} version 2022.9.5 as the main library. Furthermore, mendeleev \cite{mendeleev2014} version 0.12.1 is used to calculate some atomic properties. A Jupyter notebook, containing the code and explanations, is contained in the Supplementary Information of this paper.

For comparison, we use two other prediction methods. One are hierarchically ordered spherical environment (HOSE) codes \cite{BREMSER1978355}, a long-established method that describes atoms and their environments as strings. With those, the chemical shifts of other, similar atoms are looked up and used for prediction. From the point of view of machine learning, this could be called a nearest neighbour search. The HOSE code implementation used is a port of the HOSE code implementation of the Chemistry Development Kit (CDK) \cite{pmid29086040} and available at~\,\cite{hose}\,. This produces standard HOSE codes, not the stereo-enhanced HOSE codes of \,\cite{pmid31459832}\,.

We use the model from \cite{pmid31388784} as a modern machine learning model, which we call the ``2019 model'' in this paper. This uses a convolutional graphical neural network to combine feature vectors for an atom with those of its neighbours to do the prediction.
Evaluation of the methods was generally done using a 75:25 training-to-test split. We have decided against a separate validation set, due to the small size of the datasets. 

All data was taken from nmrshiftdb2, an open NMR database \cite{NMRSHIFTDB2}. It contains lists of chemical shift values as well as raw data of various 1D and 2D NMR experiments for a number of different nuclei. We focus on particular subsets here, as explained in the subsections of Section~\ref{sec:results}. It should be noted that the datasets we used consist of random selections of structures. It might be possible to optimize the training process with small datasets by ensuring structural diversity or even distribution in chemical space. We did not follow this and assumed the random distribution of data. In particular, we include all experimental data from nmrshitdb2 (if they fit the subsets used in Section~\ref{sec:results}). This is opposed to other work, e.g. \cite{pmid31388784} where the choice is restricted to molecules with only common elements. This also explains slightly different results using the 2019 model with data from nmrshiftdb2, apart from changes to the database over time.

For comparing the performance of the models in various conditions we report three values: The mean absolute error (MAE), the root mean squared error (RMSE), and the standard deviation $\sigma$ of the error. The standard deviation is calculated over all of the predictions of the model and is used to measure the amount of variation of the error from the mean.

\section{Results}
\label{sec:results}

\subsection{Overall behaviour}
\label{sec:overall}

\begin{figure}[]
\centering
\begin{tikzpicture}
    \pgfplotsset{set layers}
    \begin{axis}[
		xlabel=Number of spectra,
		ylabel=ppm \ref{CNN2}2019 \ref{GNN2}2023 \ref{hose2}HOSE,
		xticklabels={100,250,500,1000,2500,5000,10000,25000,44370},
		xtick={0,1,2,3,4,5,6,7,8},
        x tick label style={rotate=90,anchor=east},
        ymin=0, ymax=75]
	\addplot[color=red,mark=x] coordinates {
     (0,70.31)
     (1,64.84)
     (2,61.64)
     (3,57.77)
     (4,31.81)
     (5,3.65)
     (6,2.40)
     (7,2.11)
     (8,1.82)

};
	\label{CNN2}
	\end{axis}

    \begin{axis}[ymin=0, ymax=75,
    xticklabels={}]
	\addplot[color=blue,mark=x] coordinates {
     (0,24.80)
     (1,21.45)
     (2,22.77)
     (3,17.11)
     (4,15.00)
     (5,11.18)
     (6,9.63)
     (7,8.21)
     (8,7.65)
	};
	\label{GNN2}
	\end{axis}

    \begin{axis}[ymin=0, ymax=75,
    xticklabels={}]
	\addplot[color=green,mark=x] coordinates {
     (0,20.81)
     (1,18.99)
     (2,17.68)
     (3,18.85)
     (4,16.14)
     (5,15.2)
     (6,14.02)
     (7,12.14)
     (8,10.98)
	};
	\label{hose2}
	\end{axis}
\end{tikzpicture}
\caption{MAE of a $^{13}C$ NMR shift prediction, using increasing number of samples.}
\label{fig:13c}
\end{figure}
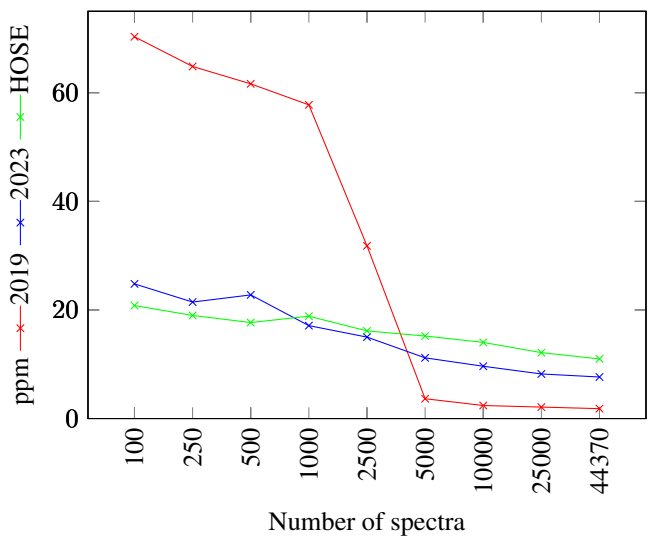

First, we wanted to compare the new 2023 model to HOSE codes and the 2019 model. In order to do so, we analyzed the predictive performance of the different models when trained on an increasing number of molecules. The results are shown in Figure~\ref{fig:13c} and Table~\ref{table:13c}. It can be clearly seen that the new model outperforms the 2019 model when trained on up to 2500 data points (structures), whereas the 2019 model performs better from 5000 data points onward. The sharp improvement of the 2019 model in that range was already seen in \cite{9842465}, and an explanation of that spontaneous improvement is yet outstanding.

The HOSE codes offer good predictive power that was previously observed in other published work, however, in some cases, a prediction based on HOSE codes is not possible, as noted in Table~\ref{table:13c}. This can happen if no examples with high enough similarity (i.e. at least one sphere) exist in the training set. The table also shows that the standard deviation of the 2019 model's results is significantly lower than with HOSE codes, indicating that the model is more stable than the HOSE code prediction.

\begin{table*}[t]
\centering
\begin{tabular}{ |c|c||c|c|c|c|c|c|c|c|c| } 
\hline
& & 100 & 250 & 500 & 1000 & 2500 & 5000 & 10000 & 25000 & 44370 \\
\hline
\hline
\makecell{2019} & MAE (ppm) & 70.31 & 64.84 & 61.64 & 57.77 & 31.81 & 3.65 & 2.40 & 2.11 & 1.82 \\
model& RMSE (ppm)  & 83.30 & 79.86 & 76.36 & 70.71 & 36.27 & 5.41 & 3.35 & 4.09 &  3.13 \\
& $\sigma$ (ppm)  & 52.29 & 52.51 & 52.69 & 48.15 & 28.05 & 6.64 & 5.08 & 5.13 & 4.57 \\
\hline
\makecell{2023} & MAE (ppm)  & 24.8 & 21.45 & 22.77 & 17.11 & 15.0 & 11.18 & 9.63 & 8.21 & 7.65 \\
model& RMSE (ppm) & 46.65 & 40.82 & 44.66 & 45.27 & 40.27 & 32.82 & 29.01 & 25.63 & 24.53 \\
& $\sigma$ (ppm)  & 45.29 & 40.44 & 44.41 & 45.13 & 40.11 & 32.77 & 28.95 & 25.58 & 24.48 \\
\hline
\makecell{HOSE} & MAE (ppm)  & 20.81 & 18.99 & 17.68 & 18.85& 16.14 & 15.2 & 14.02 & 12.14 & 10.98 \\
code & RMSE (ppm) & 35.29 & 33.44 & 30.94 & 30.32 & 30.03& 29.19& 27.95& 25.84& 24.60  \\
& $\sigma$ (ppm) & 34.72 & 33.26 & 30.74 & 30.29 & 30.01& 29.18& 27.95& 25.83& 24.5 \\
& \makecell{Missing\\predictions} & 17.6 & 19.6 & 26.3 &33.9 & 41.4 & 51.0& 57.9& 76.2 &85.2  \\
\hline
\end{tabular}
\caption{Prediction results for $^{13}C$ shifts using increasing numbers of spectra}
\label{table:13c}
\end{table*}



\subsection{Heteronuclei}

$^{13}C$ and $^{1}H$ are the most popular nuclei for NMR spectroscopy, mainly due to the natural abundance of magnetically susceptible isotopes and their presence in organic compounds. Other nuclei are also used for certain applications, but the amount of data available is much smaller. Therefore, they are a good test case for our model, where we use $^{19}F$ spectra as an example. In nmrshiftdb2, there are currently 957 structures with measured $^{19}F$ spectra. We disregard the spectra that are calculated via ab-inito calculations in nmrshiftdb2 and use only one spectrum per compound in the rare case that several spectra are recorded.

We use the same machine learning models and HOSE codes as for $^{13}C$ in Section~\ref{sec:overall}. It might be possible to improve the prediction by optimizing a model specifically for a nucleus, but this is not within the scope of this work. Generally, $^{19}F$ should behave similar to $^{13}C$ and $^{1}H$, which might not be the case e.g. for metals.

\begin{table*}[t]
\centering
\begin{tabular}{ |c|c||c|c|c|c| } 
\hline
& & 100 & 250 & 500 & 957 \\
\hline
\hline
2019 model & MAE (ppm) & 79.43 & 72.68 & 69.65 & 57.82 \\
& RMSE (ppm) & 82.54 & 84.04 & 73.23 & 61.86 \\
& $\sigma$ (ppm) & 47.32 & 93.13 & 43.18 & 41.61 \\
\hline
2023 model & MAE (ppm) & 22.25 & 15.94 & 13.32 & 9.77 \\
& RMSE (ppm) & 45.19 & 38.46 &  34.02 & 27.95 \\
& $\sigma$ (ppm) & 43.57 & 37.56 & 33.44 & 27.73 \\
\hline
HOSE code & MAE (ppm) & 12.21 & 10.53 & 7.87 & 7.38 \\
& RMSE (ppm) & 25.97 & 29.68 & 20.16 & 23.33 \\
& $\sigma$ (ppm) & 25.70 & 29.45 & 20.13 & 23.32 \\
& \makecell{Missing predictions} & 1.93 & 2.88 & 7.38 & 4.75\\
\hline
\end{tabular}
\caption{Prediction results for $^{19}F$ shifts using increasing numbers of spectra}
\label{table:19f}
\end{table*}

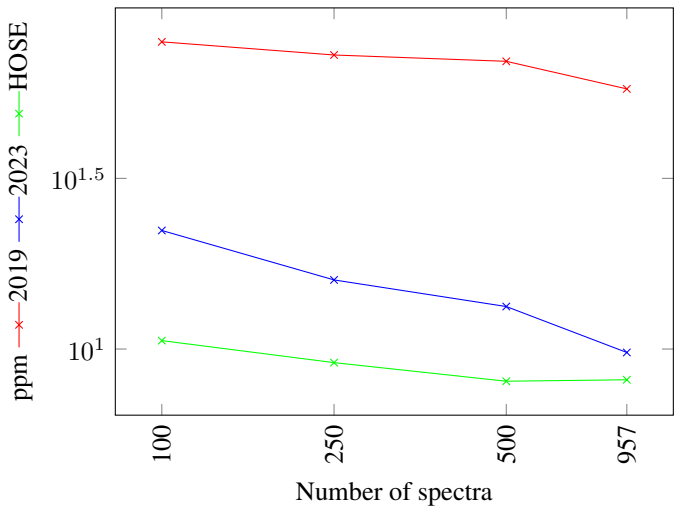
\begin{figure}[]
\centering
\begin{tikzpicture}
    \pgfplotsset{set layers}
    \begin{axis}[
        ymode=log,
		xlabel=Number of spectra,
		ylabel=ppm \ref{CNN1}2019 \ref{GNN1}2023 \ref{hose1}HOSE,
		xticklabels={100,250,500,957},
		xtick={1,2,3,3.7},
        x tick label style={rotate=90,anchor=east}]
	\addplot[color=red,mark=x] coordinates {
    (1,79.43)
    (2,72.68)
    (3,69.65)
    (3.7,57.82)

};
\label{CNN1}

	\addplot[color=blue,mark=x] coordinates {
     (1,22.25)
     (2,15.94)
     (3,13.32)
     (3.7,9.77)
	};
     \label{GNN1}

	\addplot[color=green,mark=x] coordinates {
     (1,10.584191)
     (2,9.123936)
     (3,8.047996)
     (3.7,8.127422)
	};
     \label{hose1}
	\end{axis}
\end{tikzpicture}
\caption{MAE of a $^{19}F$ NMR shift prediction, using increasing number of samples, on a logarithmic scale.}
\label{fig:19f}
\end{figure}

\begin{figure}
  \centering
  \includegraphics[width=10cm]{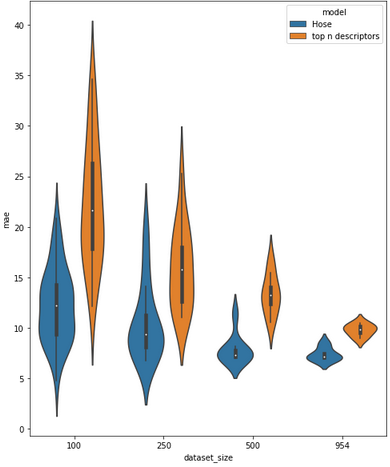}
  \caption{Comparison of the accuracies and their distribution of the HOSE code and GNN prediction.}
  \label{fig:gnnvshose}
\end{figure}

Table~\ref{table:19f} and Figure~\ref{fig:19f} show the results of the predictions based on the 957 $^{19}F$ spectra in nmrshiftdb2, using 100, 250, 500, and 957 (all) spectra for training the models. The HOSE code and 2019 model results are similar to those in \cite{9842465} (differences  are due to an older version of nmrshiftdb2 used in the paper), with the HOSE codes being significantly better than the 2019 model for those small amounts of data. We expect the model to improve with more data, similar to $^{13}C$, however, the amount of data is limited for this nucleus. Our new model, shown in blue in Figure~\ref{fig:19f}, outperforms the 2019 model when trained on 100 spectra and improves significantly, almost reaching the quality of the HOSE code model when trained on 957 spectra (note Figure~\ref{fig:19f} uses a logarithmic scale). The standard deviation of the 2023 model surpasses that of HOSE codes with 957 spectra, giving a more stable prediction here. Figure~\ref{fig:gnnvshose} shows the distribution of the 2019 model and HOSE codes, showing the significant improvement with more data.

Predictions based on HOSE codes are fairly accurate but have the inherent disadvantage that they might not give a prediction at all, as discussed previously. Machine learning models will always predict chemical shifts even for less similar molecules, as they are able to generalize. Therefore, the category ``Missing Predictions'' is only used for the HOSE codes. 

\subsection{Solvents}

The solvent used is one of the major factors influencing the chemical shift values of a particular compound due to its influence on the chemical environment of the molecule, the possibility of forming hydrogen bonds, changes in the charge state of the investigated molecule, and more. For prediction purposes, it is common practice to ignore the solvent (e.g. \cite{pmid31388784} or \cite{doi:10.1021/acs.jcim.0c00195}). More accurate predictions would require using solvent information. One problem with this is the relatively low number of spectra for particular solvents, even for $^{13}C$ and $^{1}H$ spectra. For example, nmrshiftdb2 currently has 2324 $^{13}C$ NMR spectra in Chloroform-D1, 456 spectra in Dimethylsulphoxide-D6, and 351 spectra in Methanol-D4 (those being the most common solvents in the database).

We are using those data to train separate models for each solvent and compare the results to the values achieved by using all $^{13}C$ spectra. The results are shown in Table~\ref{table:solvents}. It should be noted that the models are the same as used for the previous prediction with all solvents and the $^{19}F$ nuclei and are not optimized for a solvent-specific prediction. We can still make the following observations:

\begin{itemize}
    \item The solvent-specific training produces much better results compared to the overall model. For example, for Chloroform-D1, the 2019 model and the 2023 model reach an MAE of 24.06 respectively 4.55 ppm with 2324 spectra, whereas with all solvents the MAE is 31.81 respectively 14.60 with 2500 spectra.
    \item The overall tendency is similar to what we have seen before: The predictive quality of the 2019 model starts off with high errors and significantly improves beyond 1000 spectra. The 2023 model outperforms the 2019 model on smaller datasets, due to its quick improvements when trained on up to 2500 spectra. HOSE codes are generally doing well, but do not improve much.
    \item The 2023 model achieves errors of less than 5 ppm with 1000 spectra for Chloroform-D1. That is better than the 2023 model with all data. For the 2019 model to become better than 5 ppm, almost 5000 spectra are needed. This means that, given the available data, our new model outperforms the 2019 model. For Dimethylsulphoxide-D6 and Methanol-D4, our new model achieves reasonably good results when trained on only 456, or respectively 351 spectra. Those results are much better than the 2019 model trained on the same number of spectra.
\end{itemize}

\begin{table*}[h!]
\footnotesize
\centering
\begin{tabular}{ |c|c|c||c|c|c|c|c| } 
\hline
& & & 100 & 250 & 500 & 1000 & 2324 \\
\hline
\hline
Chloroform-D1 & \makecell{2019} & MAE (ppm) & 64.69 & 58.24 & 58.62 & 50.91 & 24.06  \\
(CDCl3) & model& RMSE (ppm)  & 80.32 & 74.71 & 74.63 & 66.18 & 34.38\\
& & $\sigma$ (ppm)  & 53.17 & 53.02 & 52.01 & 48.61 & 29.47 \\
\hline
& \makecell{2023} & MAE (ppm)  &18.89 & 7.47 & 5.23& 4.32 & 4.12\\
& model& RMSE (ppm) & 34.76 & 15.69 & 12.42 & 10.62 &  10.53 \\
& & $\sigma$ (ppm) &31.99 &15.02 &12.17 &10.47 &10.48  \\
\hline
& \makecell{HOSE} & MAE (ppm) &5.35 & 4.81 &4.32 &4.03 &3.19 \\
& code & RMSE (ppm) & 8.79 & 8.53 & 8.03 & 7.76 & 6.99  \\
& & $\sigma$ (ppm) &8.77 & 8.51 & 8.02 & 7.76& 6.99   \\
& & \makecell{Missing\\predictions} &11.92 & 14.23 & 17.30 & 19.42 & 30.25  \\
\hline
\hline
Dimethylsulphoxide-D6 & \makecell{2019} & MAE (ppm) & 91.67 & 93.09 & 85.84 (456) & n/a & n/a  \\
(DMSO-D6, C2D6SO) & model& RMSE (ppm)  & 100.41 & 100.92 & 93.76 (456) & n/a & n/a \\
& & $\sigma$ (ppm)  & 46.28 & 44.75 & 45.01 (456) &  n/a  &  n/a  \\
\hline
& \makecell{2023} & MAE (ppm)  &24.03 & 5.82 & 5.75(456) &  n/a & n/a \\
& model& RMSE (ppm) & 37.35 & 10.50 & 8.85 (456) & n/a  &  n/a  \\
& & $\sigma$ (ppm) & 31.26&  9.92& 7.61(456) &  n/a & n/a   \\
\hline
& \makecell{HOSE} & MAE (ppm) & 	4.96 & 4.27 & 3.65 (456) & n/a  &  n/a \\
& code & RMSE (ppm) & 7.61 & 6.88& 6.21(456) &  n/a &  n/a  \\
& & $\sigma$ (ppm) & 7.60& 6.87& 6.21(456) &  n/a &  n/a   \\
& & \makecell{Missing\\predictions}  & 9.95& 12.0& 10.0(456) &  n/a &  n/a  \\
\hline
\hline
Methanol-D4 & \makecell{2019} & MAE (ppm) & 78.60 & 71.92 & 69.41 (351) & n/a & n/a   \\
(CD3OD) & model& RMSE (ppm)  & 89.66 & 84.17 & 82.03 (351)  &  n/a &  n/a  \\
& & $\sigma$ (ppm)  & 49.35 & 49.95 & 49.75 (351) &  n/a  &  n/a  \\
\hline
& \makecell{2023} & MAE (ppm)  & 18.65 & 7.10 & 5.53 (351) &  n/a & n/a \\
& model& RMSE (ppm) & 32.69 & 15.31 & 9.94 (351) & n/a  &  n/a  \\
& & $\sigma$ (ppm) & 29.10& 14.53 & 9.11 (351) &  n/a & n/a   \\
\hline
& \makecell{HOSE} & MAE (ppm) & 4.67 & 4.24 & 3.64 (351) & n/a  &  n/a \\
& code & RMSE (ppm) & 8.15 & 8.05& 7.37 (351) &  n/a &  n/a  \\
& & $\sigma$ (ppm) & 8.13& 8.04& 7.37 (351) &  n/a &  n/a   \\
& & \makecell{Missing\\predictions}  & 8.42& 6.75&12.5 (351) &  n/a &  n/a  \\
\hline
\hline
\end{tabular}
\caption{Prediction results for $^{13}C$ shifts using increasing numbers of spectra. n/a indicates that not enough data were available, numbers in brackets indicate number of compounds used, deviating from the top header.}
\label{table:solvents}
\end{table*}

\begin{figure}[]
\centering
\begin{tikzpicture}
    \pgfplotsset{set layers}
    \begin{axis}[
		xlabel=\makecell{Number of spectra\\  \ref{CNN3}2019 all \ref{CNN4}2019 CDCl3 \ref{GNN3}2023 all \ref{GNN4}2023 CDCl3},
		ylabel=ppm,
		xticklabels={100,250,500,1000,\makecell{2324/\\2500}},
		xtick={0,1,2,3,4},
        x tick label style={rotate=90,anchor=east},
        ymin=0, ymax=75]
	\addplot[color=red,mark=x] coordinates {
     (0,70.31)
     (1,64.84)
     (2,61.64)
     (3,57.77)
     (4,31.81)

};
	\label{CNN3}
	\end{axis}

    \begin{axis}[ymin=0, ymax=75,
    xticklabels={}]
	\addplot[color=red, dotted,mark=*] coordinates {
     (0,64.69)
     (1,58.24)
     (2,58.62)
     (3,50.91)
     (4,24.06)
	};
	\label{CNN4}
	\end{axis}

    \begin{axis}[ymin=0, ymax=75,
    xticklabels={}]
	\addplot[color=blue,mark=x] coordinates {
     (0,57.85)
     (1,31.54)
     (2,22.95)
     (3,17.78)
     (4,14.60)
	};
	\label{GNN3}
	\end{axis}

    \begin{axis}[ymin=0, ymax=75,
    xticklabels={}]
	\addplot[color=blue,dotted,mark=*] coordinates {
     (0,18.89)
     (1,7.47)
     (2,5.23)
     (3,4.31)
     (4,4.12)
	};
	\label{GNN4}
	\end{axis}

\end{tikzpicture}
\caption{MAE of a $^{13}C$ NMR shift prediction, using increasing number of samples.}
\label{fig:solvents}
\end{figure}
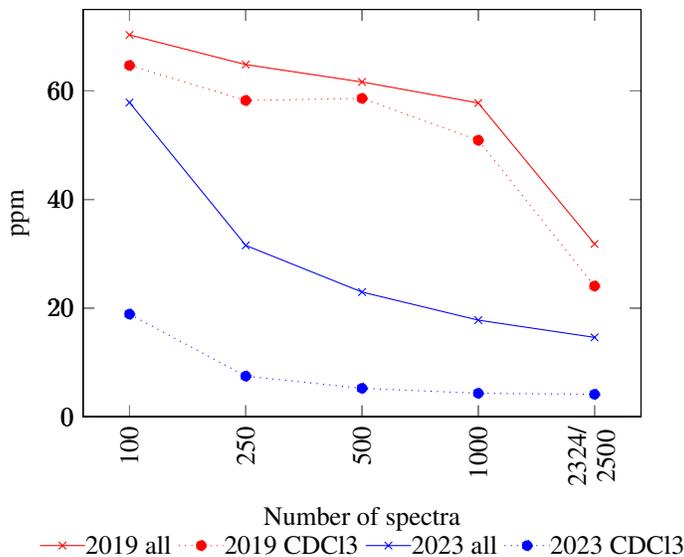

Figure~\ref{fig:solvents} shows the MAEs achieved by the two models trained with all data and CDCl3 only. It is clearly visible that the 2023 model outperforms the 2019 model. Furthermore, the CDCl3 predictions are not only better with each model than the predictions with all data, but the improvement with the 2023 model is higher than with the 2019 model.

To further verify our results, we have predicted the shifts of only CDCl3 spectra, but with all data used for training, using the 2019 model. For this, we have divided the CDCl3 spectra into four equal parts and trained models with all non-CDCl3 data and three of those parts. The fourth part was then used as a test set, predicting all shifts of it. The average errors of those for test sets were: MAE 2.04, RMSE 2.65, and $\delta$ 2.60. This confirms that the 2019 model is able to achieve good results also on the CDCl3 data alone, given enough data and that the good results of the 2023 model with CDCl3 data are not due to those data.

\begin{figure}
  \centering
  \includegraphics[width=15cm]{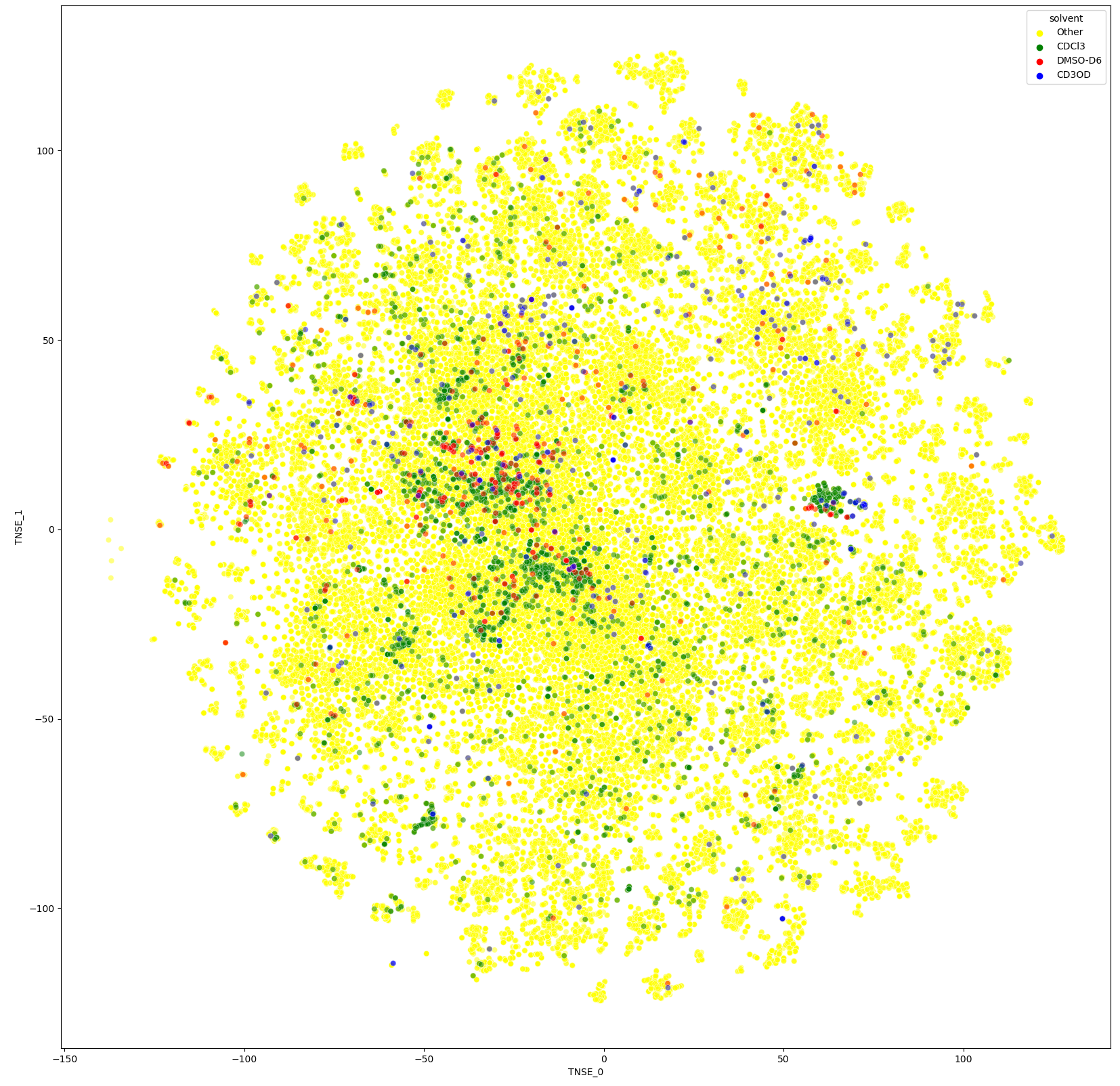}
  \caption{A plot of the compounds of nmrshiftdb2, distinguished by solvent, in chemical space. The calculation uses Extended Connectivity (ECFP) fingerprints to calculate descriptors and t-distributed stochastic neighbor embedding (t-SNE) for dimension reduction. The two major components are plotted. Using code from \cite{space}\,.}
  \label{fig:chemspace}
\end{figure}

In order to verify that the distribution of the compounds is not dependent on solvents, we have plotted the compounds in a chemical space chart in Figure~\ref{fig:chemspace}. Here, all three solvents show a distribution similar to the overall database. It should be noted that all methods would be equally affected by any potential distortions.

%

\section{Discussion}

In this paper, we have designed a new model based on message-passing graph neural networks for predicting chemical shifts. The new network is intended, in contrast to existing models, to work with small amounts of training data.

Testing this new model with $^{19}F$ data shows that it is possible to decrease the error rates significantly below the values of a standard deep learning model. Specifically, when trained on all $^{19}F$ data from nmrshiftdb2, our model achieves an MAE of 9.95 ppm, whereas the standard deep learning model only achieves an MAE of 57.82 ppm. In a similar fashion, it is also possible to improve the predictions on $^{13}C$ chemical shifts with a particular solvent. With the new model, we get an MAE of 4.5 ppm for all spectra, whereas the standard model only achieves 24 ppm. This clearly shows our model performs significantly better on smaller datasets.

Analysing Tables~\ref{table:13c} and \ref{table:19f}, we can conclude that good results can be achieved by training the new model on roughly 500 structures. This result is empirical and might depend on e.g. the diversity of the structures, but since it holds true for $^{19}F$ and $^{13}C$ it can be considered a rough threshold value.

Our model is only optimized for the prediction of $^{19}F$ chemical shift data and was used as-is for the other test cases. This means, that a more specialized model might perform better still for e.g. a particular nucleus or solvent. One way to improve performance would be to adjust the feature selection specifically for a dataset. The approach of using one model should work within solution NMR of organic compounds, where it could be useful to train a model specifically for a certain compound class. On the other hand, there are areas where this is unlikely to work, e.g. when predicting inorganic compounds or solids. The overall approach however could still prove useful as the availability of data is a problem often faced in research.

In this work, we have not tested all currently available models with different amounts of data. Some of them were published after the 2019 model and claim to have slightly better results using large datasets. Therefore, it is possible that they do better with small datasets, however, we expect the differences to the 2019 model to be negligible, as they have not been specifically built for and tested on smaller datasets.

\section{Conclusion}

We have introduced a novel machine learning model which is able to achieve reasonable NMR shift prediction results with a low number of samples. On $^{13}C$ NMR spectra, the model achieves an MAE of less than 5 ppm when trained on 1000 data points. Tests were performed on $^{19}F$ shifts and solvent-specific $^{13}C$ shifts. In all cases, our model outperforms an existing state-of-the-art machine learning model. Further improvements might be possible by optimizing aspects of the model for specific datasets, but in this work, we wanted to focus on the performance boost even the generalized model achieves. This approach might be expanded into other areas such as inorganic compounds, but there further adjustments are likely needed.


\section*{Acknowledgements}

The authors thank all participants in the UT module ``Machine learning''  working on this problem: Karl Kristjan Kaup, Artur Kurvits, Ellen Leib, Dmytro Pashchenko, Kyrylo Riazantsev, and Daniel W{\"u}nsch as well as Holger A. Scheidt (Leipzig University) for help with the manuscript. S.K acknowledges funding by De Montfort University for computational facilities (VC2020 new staff L SL 2020).

\section*{Supplemental Info}

The code of the project, together with explanations, is provided as a Jupyter notebook at https://colab.research.google.com/drive/1yKTRjpWzR8T199eCokuJfd9Y5o2oNtPp.

\bibliography{achemso-demo}

\end{document}